# Characterization and performances of new indium loaded organic liquid scintillators, based on novel indium carboxilate compounds.

I. Barabanov[1], L. Bezrukov[1], C. Cattadori[2], N.Danilov[3], A. Di Vacri[4], N. Ferrari[4*], V. Kornoukhov[1], Y.S. Krylov[3], N. Nesterova[5], S. Nisi[3], E.Yanovich[1].

[1]Institute for Nuclear Researches – RAS; Moscow (Russia)
[2]Istituto Nazionale di Fisica Nucleare – Milano Bicocca; Milano (Italy)
[3]Istituto Nazionale di Fisica Nucleare – Laboratori Nazionali del Gran Sasso; L'Aquila (Italy)
[4]Institute for Physical Chemistry and Electrochemistry– RAS; Moscow (Russia)
[5]Institute of Organic Chemistry – RAS; Moscow (Russia)

**Abstract:** A novel formulation to dope organic liquid scintillators (OLS) with indium at concentrations up to 10% is presented: it is based on specific indium carboxylate compounds adequately synthesized. The produced In-OLS has been characterized: it has light yield 8500 ph/MeV at indium concentration 5.5% and light attenuation length of 2,5 m at $\lambda$=430 nm. The scintillator properties were stable during all time of investigation (~ 1 years). The produced In-OLS is compared to other In-OLS formulations and shows superior performances. The developed methodic to metal dope OLS can be easily extended to other metals as Gd, Nd, Cd.

## 1 Introduction

The development of Indium loaded scintillator (In-LS) has been proposed and initiated by R. Raghavan [1] in the '70s to observe the low energy solar neutrino from pp,$^{7}$Be and PEP reaction via neutrino capture on $^{115}$In (I.A. 96%, specific activity ) with an energy threshold of 114 keV.

$$^{115}\text{In} + \nu_e \rightarrow {}^{115}\text{Sn}^*(612\text{ keV}) + e^-$$
$$^{115}\text{Sn}^* \rightarrow {}^{115}\text{Sn} + \gamma_1(115\text{ keV}) + \gamma_2(497\text{ keV}) \qquad (1)$$

The transition caused by $\nu_e$ absorption proceeds through an excited isomeric state of $^{115}$Sn*, to the $^{115}$Sn ground state, with the emission of a two $\gamma$-rays cascade, 4.76 $\mu$s delayed. The occurrence of three events, i.e. an electron and two delayed gamma rays properly located in time and space around the electron event, is a specific $\nu_e$ capture tagging that can be used to reduce the formidable background in the 1 MeV range, generated by external neutrons and $\gamma$-rays, and internal radioactive isotopes as $^{176}$Lu,$^{138}$La,$^{232}$Th,$^{238}$U. The feasibility of experiments *a la* LENS (Low Energy Neutrino Spectroscopy), will not be discussed in this paper, but we point out that, beside challenging radiopurity issues, it strongly depends on the space, time and energy resolution of the detector,. The feasibility of LENS experiment has been seriously considered [2] [3], and it is still discussed []. The proposed detector is an array of cells having 5x5 cm$^2$ cross section by few meters length; cells filled with indium doped organic liquid scintillator (In-OLS) are intercalated with undoped OLS cells. The total In mass is ~ 20 t [1**]**.
The crucial part of the detector is the In-OLS, which should satisfy the following requirements:

- indium concentration in ranging from 5% to 10%,

[*] Deceased

- light yield (LY) not less than 8000 ph/MeV
- transparency in the 415 - 450 nm wavelength range order of few meters,
- stability of the optical and scintillation parameters along several years.

The development of such an In-OLS is a challenge because as a rule of thumb the metal concentration diluted in an OLS and its LY and transparency are anti-correlated. In the following sections we will review the formulation that we have developed to produce high quality In-OLS (Section 2), as well as their optical, light yield and $\gamma$–spectrometry performances (Section 3). Finally the In-OLS developed in this work will be compared to In-OLS developed by other research groups (Section 4) [4][5][6].

## 2 Doping Organic Liquid Scintillators with indium

Metals in form of their inorganic salts are only poorly soluble in OLS; therefore to produce In-OLS the first task is to choose an Indium compound exhibiting satisfactory solubility in organic solvents. For reasons extensively explained in [7][8][9][10], and mainly related to solubility and stability of final In compound, we chose In-Carboxilates (In-CBX) i.e. the Indium salt of carboxylic acids (CA). The development of a chemical formulation to produce In-CBX salts, free from optical impurities and from light quenching agents, easily dissolvable in OLS is the goal of our chemical work; in this section we summarize the main results.
Extensive investigations of CAs from $C_3$ to $C_9$ lead to the selection of two CAs that satisfy the solubility and stability (of optical and fluorescent properties) requirements, namely isovaleric ($C_5$) and 2-metylvaleric acids ($C_6$); their specific molecular weight and steric factor allow to produce In salts suitable for our purposes. Their physical-chemical properties have been investigated in details; the constants of ionization, distribution and dimerization in 1,2,4 try-methyl benzene (TMB) also known as pseudocumene (PC) as well as the influence of relevant parameters as the pH of the water phase, reagent concentrations, temperature etc. on the distribution of In between water and organic phases have been determined [2,3]. By reaction of CAs with InCl3 in aqueous environment, In-CBX are formed. Two methodic to drive In-CBX in organic solution were then developed:

A. traditional liquid extraction of In with CAs from water solution of In chloride
B. synthesis, by precipitation, of solid water-free In-CBX followed by dissolution in organic solvent.

PC and/or mixtures of PC with Mineral Oil (MO) ware used as organic solvent. The synthesis of In-OLS based on IniVA turned out to be possible by methodic A when adding neutral phosphor-organic compounds (NPOC). The solubility of $In(iVA)_3$ in PC is small (3.2 g/l) but addition of NPOC results in formation of the $In(iVA)_3 \cdot 3$ NPOC compound that exhibit increased solubility (120 g/l)in PC.
However the measured LY (4000-5000 ph/MeV) of In-LSs based on $In(iVA)_3$ at In concentration ~ 5-10% was not really satisfactory; moreover the strong unpleasant smell and some toxicity of iVA made it difficult produce and handle the In-OLS in large quantity. Finally, and worse, the optical transparency of In-OLS obtained by $In(iVA)_3 \cdot 3$ NPOC (several NPOC have been tested), showed a decay of the order of several months.
A major improvement came from usage of 2-metylvaleric acids (2MVA); it came out that it is possible to keep In-2MVA salt in OLS solution without adding NPOC which are quenching agents. The study of In extraction with solutions of 2MVA showed [3] that three compounds come into organic phase depending on the leading synthesis parameters:

1. $In(2MVA)_3 \cdot 3HMVA$,
2. $In(2MVA)_3$
3. $[In(2MVA)_x(OH)_{3-x}]_n$, where n-polymerization degree.

It was measured that the extracted complexes in average include 1-2 molecules of water per In atom; the water molecules coming with compounds (1) and (2) must be removed, otherwise the In-OLS loses transparency in 20 –30 days. The stability of optical properties was improved removing water from the obtained organic phase with agents as Na sulfite or A-4 zeolite. The disadvantages of In-OLS based on liquid extraction procedure were i) not high enough LY, ii) low stability and unavoidable pollution (especially by radioactive $^{40}K$) due to contact with dehydrating compounds. A major improvement came from controlling the synthesis parameters to produce through methodic B the solid compound $[In(2MVA)_x(OH)_{3-x}]_n$; we were able to identify that LY and stability of In-OLS depends on x and n coefficients, therefore we identified and properly tuned the relevant synthesis parameters to maximize LY and stability ($x \sim 0.8$, $n \sim 10$) .
The In2MVA compound of type 3 is the best for our purposes as

- polymerization results in stabilization of extracted complex and $OH^-$
- substitution of a part of $2MVA^-$ - anions (molecular mass 115) by $OH^-$ - anions (molecular mass 17) resulted in significant decrease of the included compound molecular mass i.e. in increase of organic solvent in In-OLS; as a result, LY increases.

Figure 1 shows the IR spectra of the $[In(2MVA)_x(OH)_{3x}]_n$ complex synthesized with the optimized procedure.

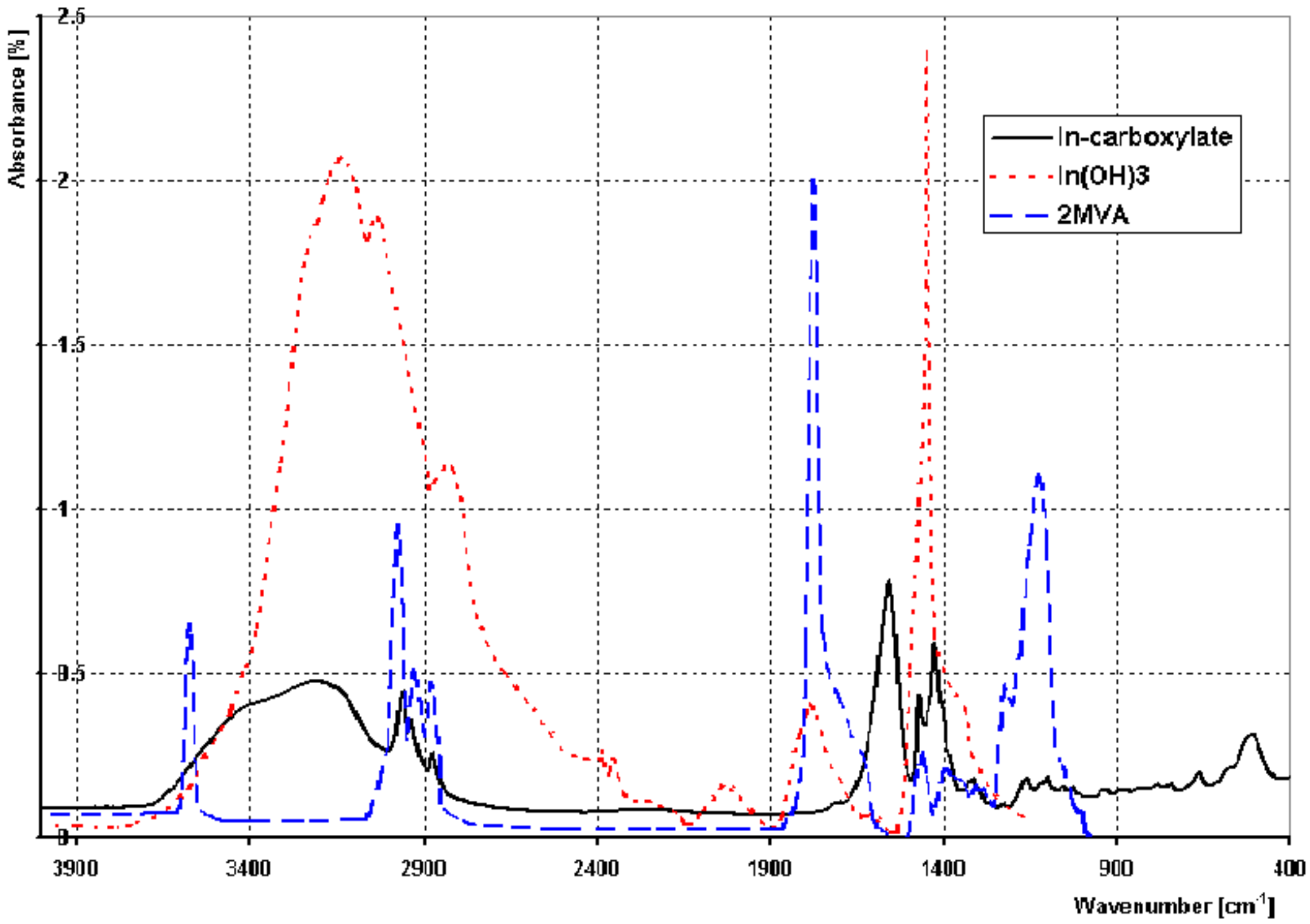

**Figure 1 IR spectra of $[In(2MVA)_x(OH)_{3-x}]_n$, In(OH)3 and 2MVA.**

## 3 Performances of In-OLS: light yield and light absorbance.

The In-OLS were characterized by two parameters relevant to be adopted in a large scale ionizing radiation detector, namely light yield (LY) and light absorbance ($A_{\lambda i}$) at the proper range of wavelengths. The absorbance of In-OLS was measured by UV/VIS spectrometer Perkin-Elmer Lambda 18 using quartz cylindrical cells of 10 cm optical length: attenuation length ($L_{\lambda i}^{att}$), defined as the length at which the intensity of light is reduced of a factor 1/e, is obtained from the measured absorbance value ($A_{\lambda i}$) according the following formula

$$L_{\lambda i}^{att}\,[cm] = d\,[cm] / (A_{\lambda i} \cdot \ln 10) \qquad (2)$$

where d is the optical length of the measured sample (i.e. length of cells) and $A_{\lambda i}$ are the absorbance values referred to the absorbance at 600 nm.

Light output of samples was determined by a standard technique, comparing pulse height amplitude (PHA) spectra collected irradiating samples with proper gamma source to PHA spectra collected irradiating the standards (BC-505 and TMB) in identical geometrical conditions with the same source. As scintillator additives both for samples and standard, we used 2-(4 biphenyl)-5-phenyloxazol (BPO) 4-5 g/l; BPO has maximum emission at 325 nm, therefore the addition of a secondary fluor as bis-MSB was not needed to match the PMT quantum efficiency, a standard bialkali photocatode.

The LY of In-OLS, given by the ratio of the position of the Compton hedges in collected energy spectra, express the In-OLS LY as percentage of yield of the standard (12000 ph/MeV). The LY of the produced In-OLS and reference were in turn measured in ~10 $cm^3$ volume cells optically coupled to a Philips …. PMT, irradiating the cells with radioactive γ-sources $^{241}$Am(60 keV), $^{57}$Co(122 keV and 136 keV), $^{22}$Na (511 keV and 1275 keV), $^{137}$Cs (662 keV) and $^{54}$Mn (835 keV) to study the In-OLS response versus the energy of the incoming radiation. The PMT signal was then processed by a standard nuclear spectroscopy electronic chain.

Finally, 2 quartz cells of ~ 2 l volume (see Figure 8), have been filled with In-OLS produced by this and by another formulation [4][5][6] to compare their γ-spectroscopy performances and stability in same experimental conditions (Section 3.3)

## 3.1 Light yield

Figure 2 shows the LY curve of our indium loaded OLS based on In-2MVA solid salt diluted at different percentages in PC. LY in the range 55%-75% of standard namely 7000-9000 ph/MeV were obtained at In concentration ranging from 50 to 100 g/l. This result is relevant and meet the specifications required for large scale detector based on In-OLS.

Figure 3 shows spectra collected when irradiating the ~10 $cm^3$ cells filled with In-OLS with different γ-ray sources; the recorded spectral shapes properly scale with energy. The full γ-absorption peak is clearly visible in the In loaded sample, thanks to the increased density and Z; its intensity is proportional to the In concentration. Figure 4 shows the obtained energy calibration obtained when irradiating the In-OLS in the 2 l quartz cell described in Figure 8. The corresponding energy resolution function is

$\sigma/E = 0.97 + 98.0 \cdot E^{-1/2}$, and it is shown in Figure 5; the energy resolution been found at 120 keV and 662 keV is 10% and 6% respectively. The small dependence of the energy resolution from the source position is due to some instrumental effect and has not been further investigated in this work

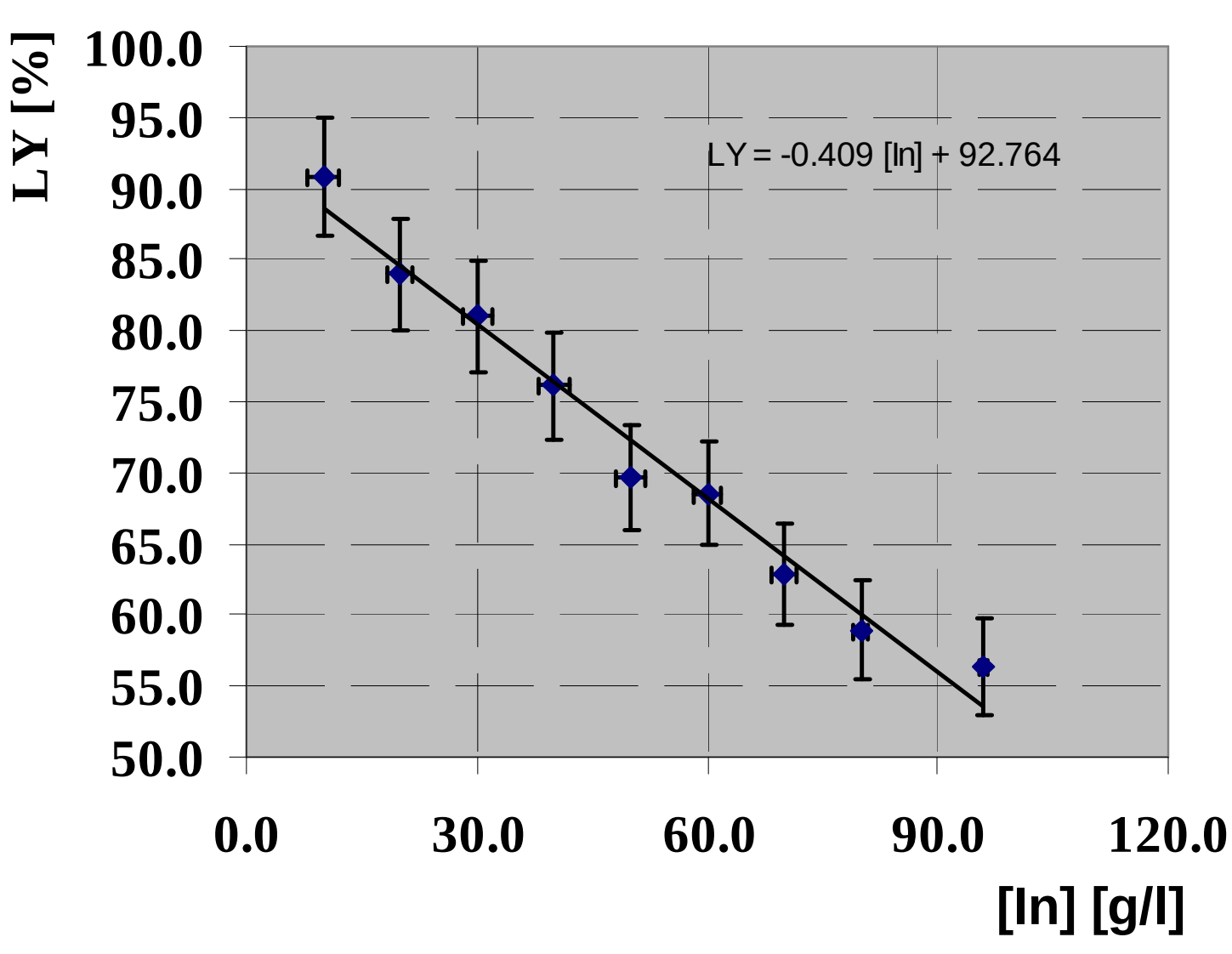

**Figure 2 LY as a function of indium concentration for our best formulation of In-OLS based on In-2MVA solid**

**salt diluted in PC. Primary fluor is BPO 5 g/l.**

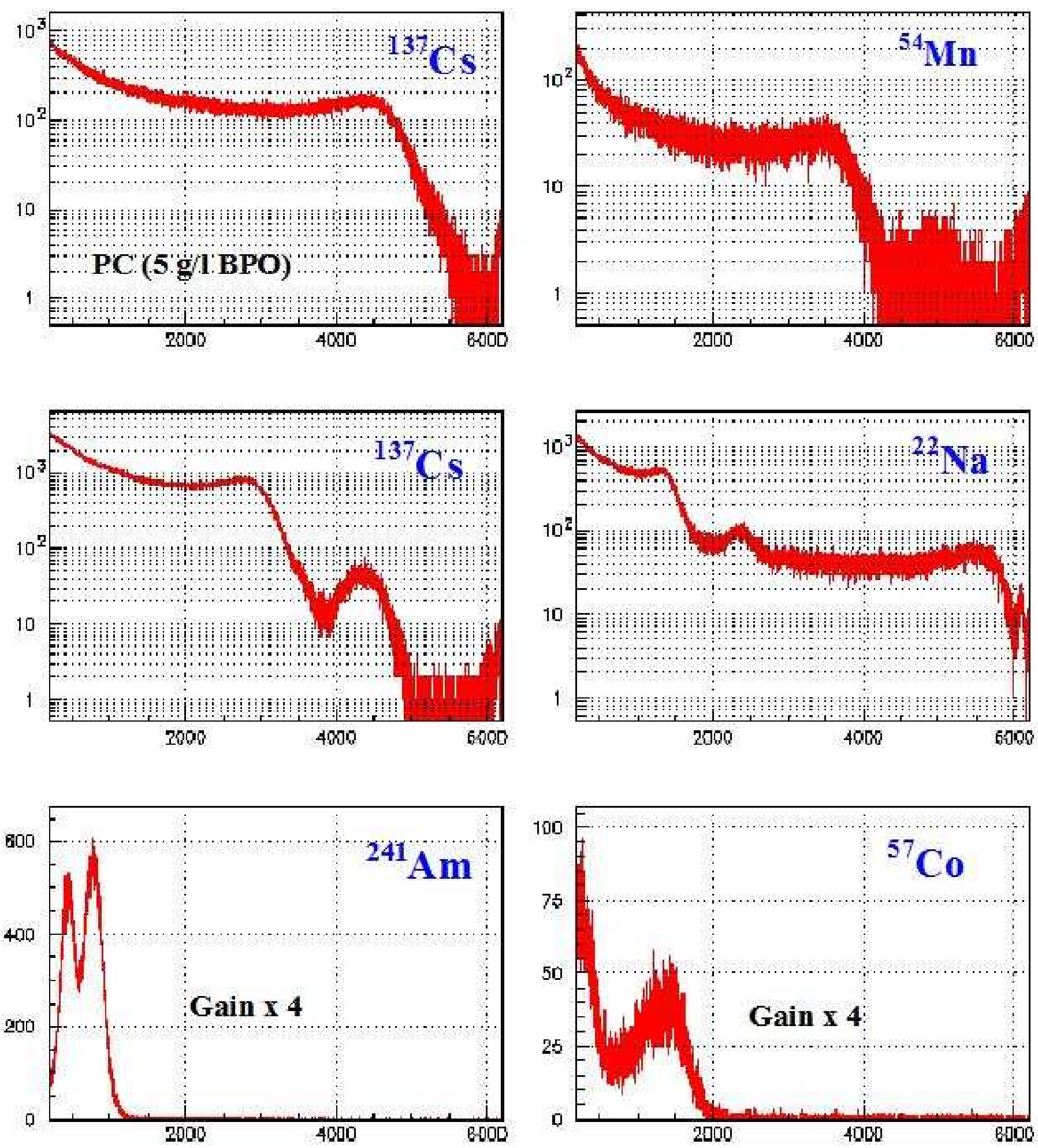

**Figure 3 PHA spectra collected irradiating the In-OLS based on In-2MVA solid salt with different γ-ray sources: $^{241}$Am(60 keV), $^{57}$Co(122 keV and 136 keV), $^{22}$Na (511 keV and 1275 keV), $^{137}$Cs (662 keV) and $^{54}$Mn (835 keV). [In]= 8 g/l, BPO = 5 g/l.**

## 3.2 Light absorption

Figure 6 shows the measured light absorbance and derived attenuation length measured in 10 cm optical length cells. The developed In-OLS has attenuation length ($L^{430}$) of 2.5 m at 430 nm for Indium concentration of 80 g/l. The dependence of the attenuation length from In concentration was also studied; the attenuation length of samples having [In] = 75 g/l and 55 g/l, respectively with and without primary fluor, was found to be 290cm, and 390 cm respectively. Moreover, as shown in Figure 7, the absorbance spectrum of the In-OLS in the UV region has the same shape of pure PC one; this demonstrate that no extra absorption band are introduced by the adopted reagents or by the In-2MVA compound. In other words no competitors (quenchers) to the energy transfer from PC to primary fluor are present. Such high light attenuation length of the final synthesis product (In-OLS) can be obtained only if adopted reagents are properly purified; the carboxylic acids were purified by double vacuum distillation, and the organic solutions were purified through $Al_2O_3$ column.

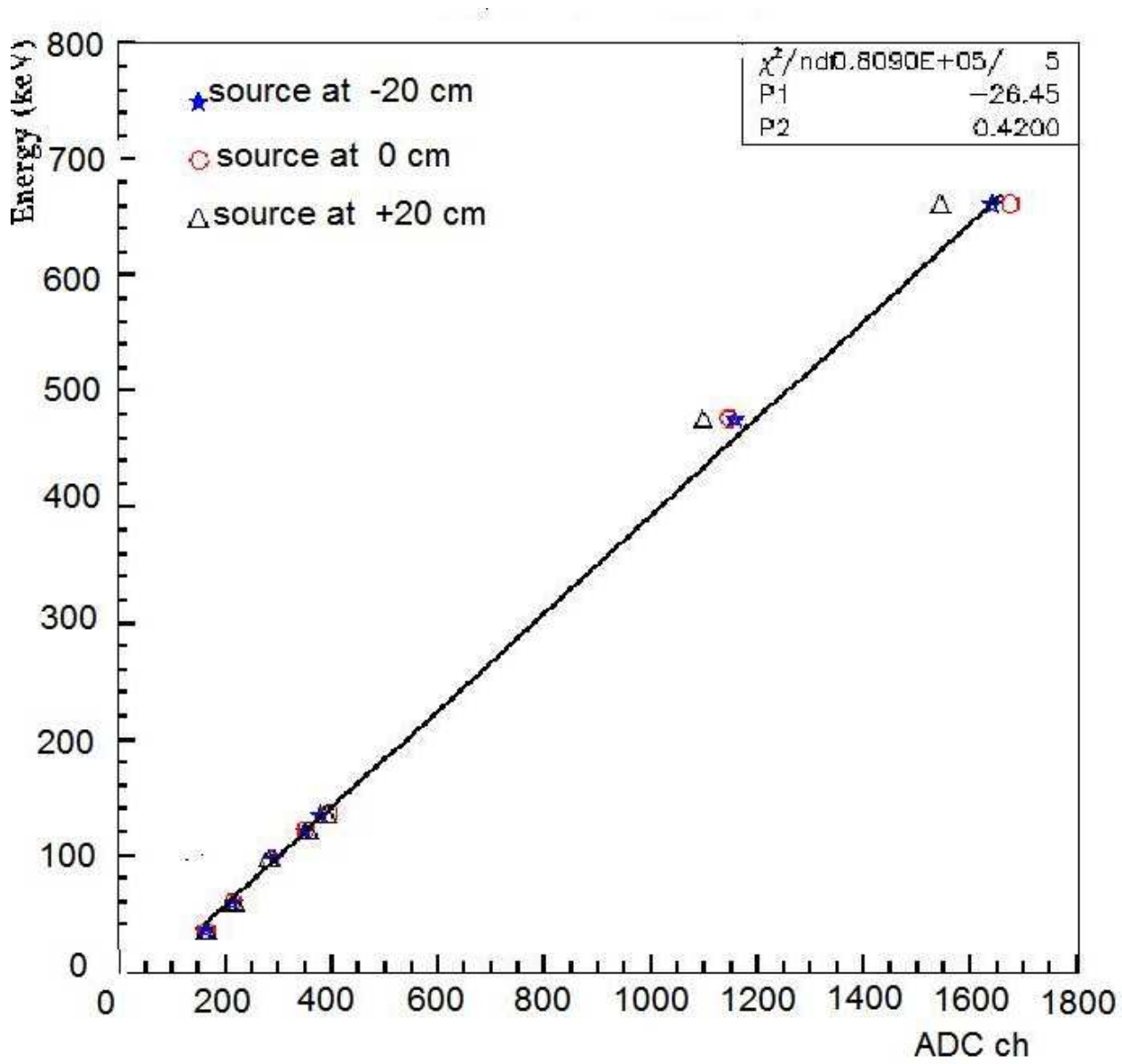

**Figure 4 Energy calibration of In-OLS based on In2MVA, irradiated in 2 l quartz cells with several gamma sources placed at different positions along the cell axis. [In] = 55 g/l, BPO = 4 g/l.**

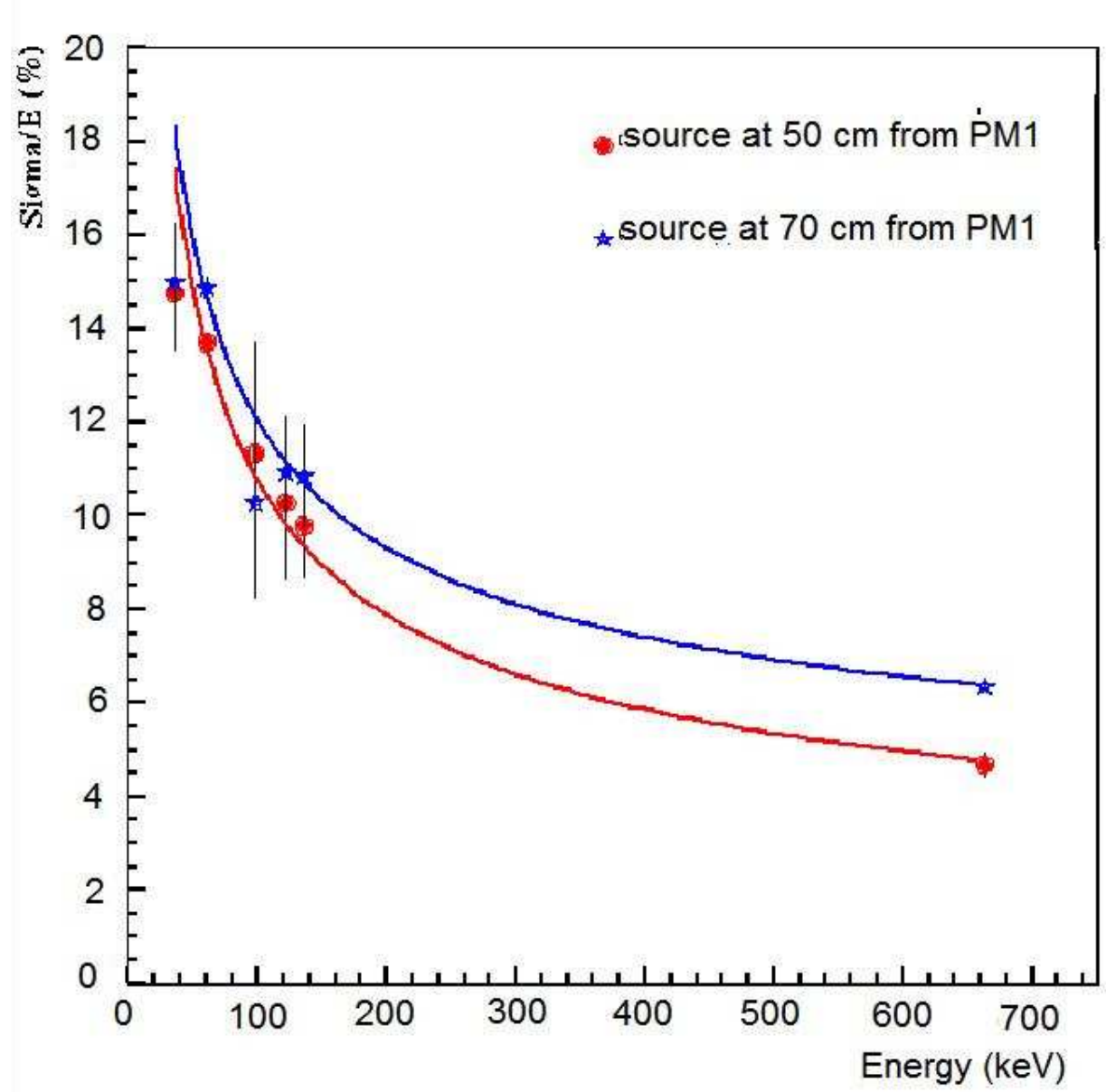

**Figure 5 Energy resolution as a function of energy for In-OLS based on In2MVA irradiated in 2 l quartz cells with γ-ray sources, placed at different positions along the cell axis. [In] = 55 g/l, BPO = 4 g/l.**

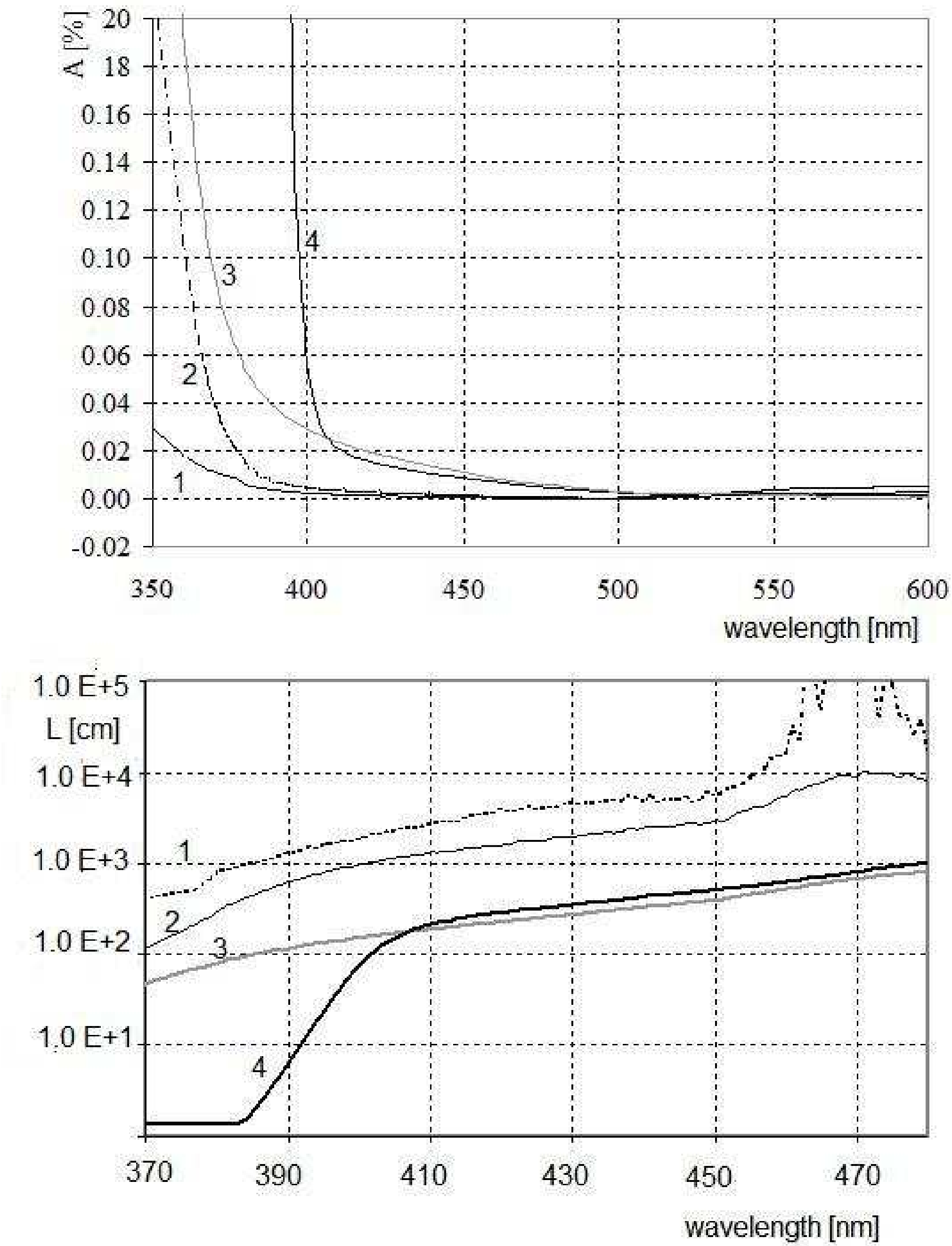

**Figure 6 Absorbance spectra (top) and derived attenuation lengths (bottom) for pure Methyl-Valeric Acid (1), pure PC(2), InLS at [In] = 80 g/L (3); InLS at [In] = 55 g/L + BPO = 5 g/L (4). Measurement is performed in cell having 10 cm optical length**

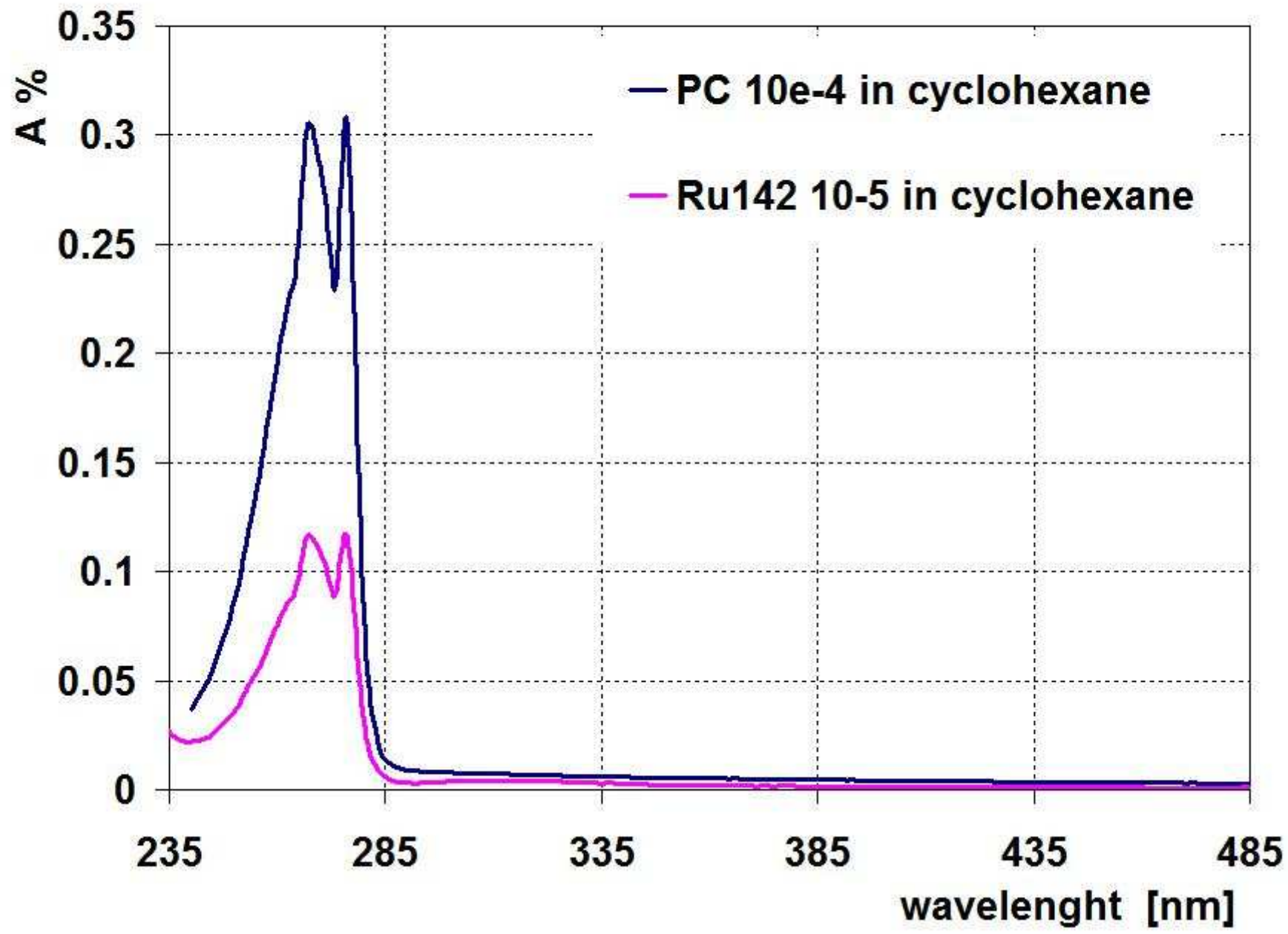

**Figure 7 Comparison of Absorbance spectra of pure PC and In- OLS based on our In-2MVA compound.**

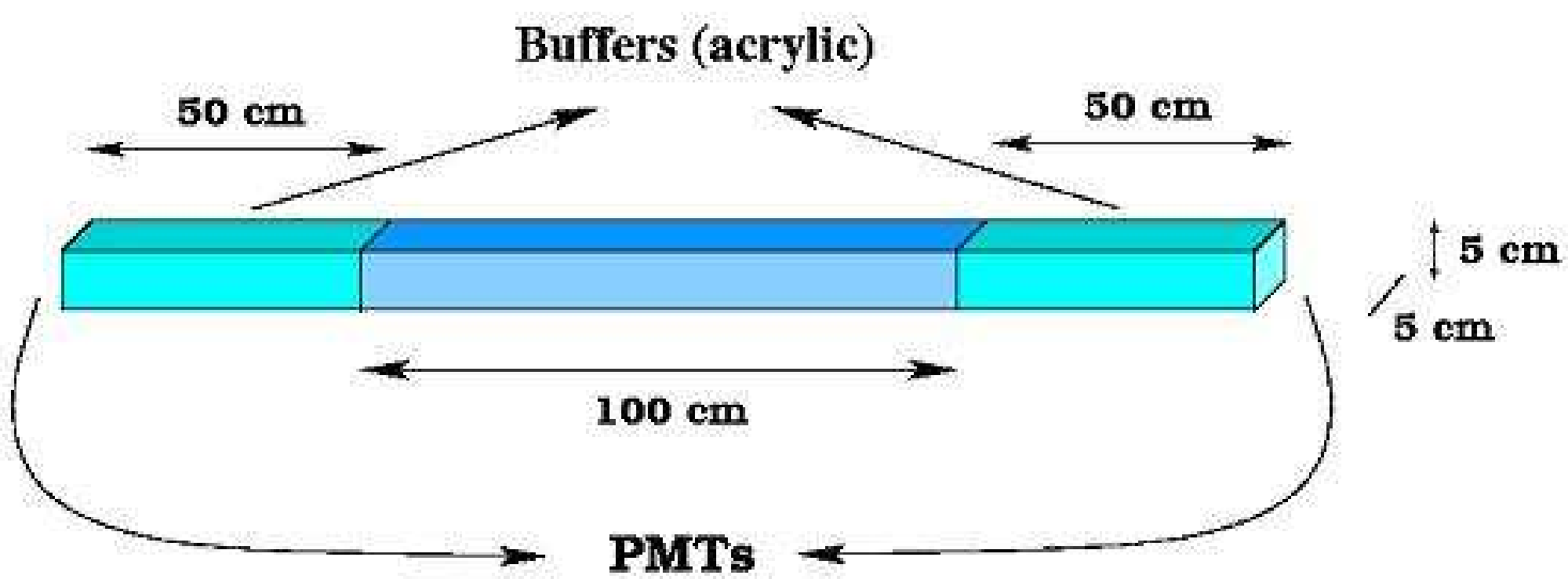

**Figure 8 Schematic of the quartz cells adopted for comparative and stability investigation of differently formulated In-OLS in a low background environment. Cells have been developed by MPIK group that has also developed the In-acac formulation to produce In-OLS based on PXE.**

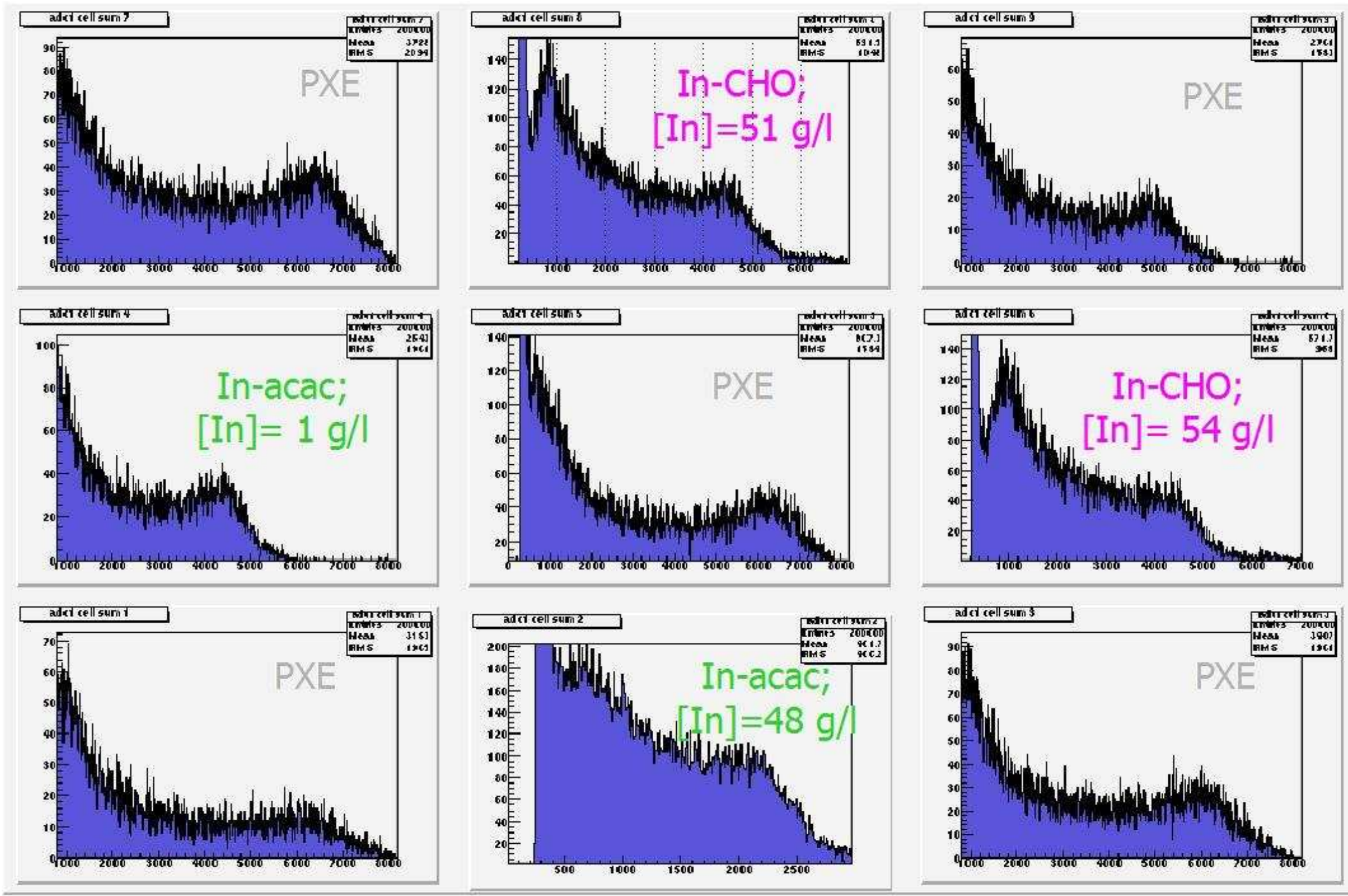

**Figure 9 Spectra collected with 9 cells 2 l volume, 5cm x 5cm x 100 cm dimensions, filled with reference unloaded OLS (PXE) and 2 different type of In-OLS. In-CHO is the object of this paper , In-acac has been developed at MPIK and reported in [4][5][6].**

## 3.3 Stability of In-OLS

The stability of In-OLS and in general of metal loaded OLS is the most critical issue for experiments whose data taking is planned to last over several years. Therefore a ~1 year survey of the In-OLS light attenuation length has been performed using two different and independent techniques. An array of 9 quartz cells developed in the frame LENS (an R&D project pursued by an U.S.A. – Russian –European at LNGS in the years 2001-2004) [2] where filled with In-OLS based on two different compounds [6][11]. The single cell schematic is shown in Figure 8; cells are quartz made, optically polished on all the faces to be operated as a light guide in total internal reflection (TIR) regime. The light generated in the scintillator by ionizing radiation is then collected by TIR at the two end-faces of the cell and finally red-out by two PMTs (one at each face). Energy spectra are built up from single PMTs signal and then analyzed for light attenuation study. After a first run when all the cells were filled with PXE to equalize the response of the different read-out chains, 4 cells where then filled with In-OLS based on two different In organic compounds (In-2MVA and In-acac) and different organic OLS (PC and PXE) and 5 cells remained filled for reference with PXE. Figure 9 shows the energy spectra collected when irradiating the cell array with $^{137}$Cs source, that could be placed at different positions. The Compton Edge (CE) structure is clearly visible in all the spectra. The position of the CE in the PHA spectra is recorded scanning the cell along its main axis. The ratio of the CE position when source is at 90 cm and at 10 cm from one of the two PMTs taken as reference, is the $R_{90/10}$ parameter; it is related to the light attenuation length and if monitored in time, while keeping all the geometrical parameters unchanged, it express the stability of the In-OLS under investigation. Figure 10 shows the plot of the $R_{90/10}$ parameter as measured during ~ 4 months. No decrease of $R_{90/10}$ therefore no decrease of the attenuation length was observed over 4 months. Moreover, we have surveyed over 6 months, by spectrofotometric measurements in 10 cm optical path cells, the In-OLS light attenuation length. No decrease of the attenuation length has been observed over 6 months (see Figure 11) . We want to emphasize that the In-OLS stability strongly depends on the adopted indium organic compound, and mainly on the compound synthesis parameters and conditions.

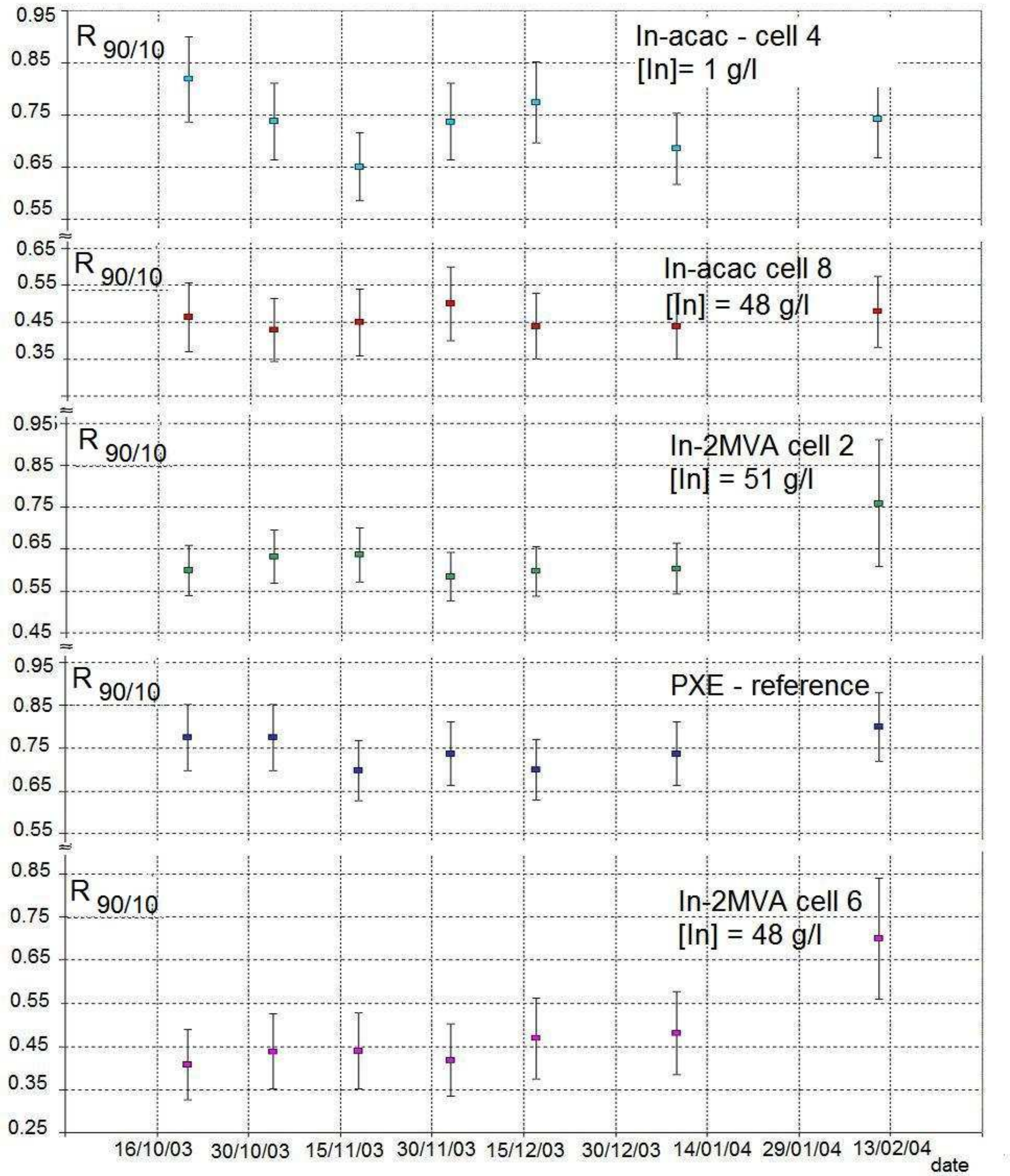

**Figure 10 Stability survey of In-OLS based on In-2MVA (cells 2 and 6) and In-acac (cell 4 and 8) compounds. For comparison one PXE filled cell is also reported. The plotted quantity is $R_{90/10}$ .**

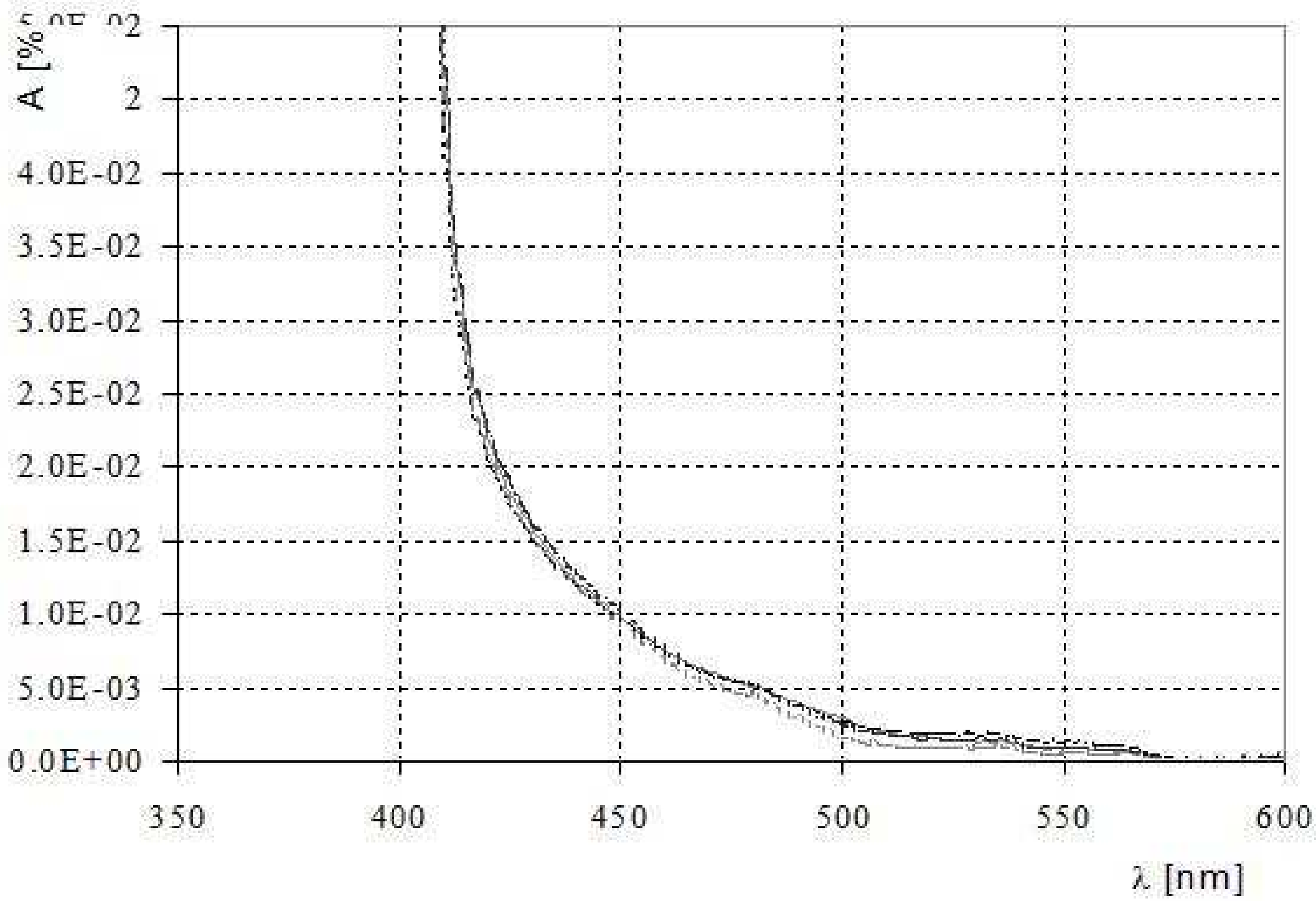

**Figure 11 Stability survey of InLS with 5 g/L of BPO and 15 mg/L of bisMSB at [In] = 55 g/L. Spectra collected over 6 months overlap perfectly. No absorbance increase is observed.**

## 4 Comparison of In-OLS developed in this work with other In-OLS formulations.

Other developments of In-OLS were done in the same period (2000-2003) and in the framework of the former LENS collaboration, that investigated the scientific relevance and feasibility of a full scale solar neutrino experiment at LNGS (see f.i.[11]). In this section we will compare the performances of the two In-OLS independently formulated [4][5]. Table 2 summarize the characteristics and the performances of In-OLS measured in the cell array. When the performances of the three cells having approximately the same indium concentration are compared, the In-OLS obtained by dilution of In-2MVA solid salt in PC shows twice the LY (95% of PXE) and 1.5 time the effective light attenuation length[1] ($L_{att}^{eff}$ = 140 cm) of In-acac based formulation, which exhibit a LY of 50% of PXE and $L_{att}^{eff}$ = 115 cm at a slightly lower indium concentration: both LY and $L_{att}^{eff}$ depends strongly on indium concentration.

[1] $L_{att}^{eff}$ is the effective attenuation length. It corresponds to the light attenuation length weighted by the PMTs photocatode quantum efficiency and by geometrical factors.

| Cell # | In LS | [In] [g/l] | Mass [kg] | LY(1) % @ 50 cm | $L_{att}^{eff}$ [cm] | LY [2] % | $L_{att}^{430}$ [3] [cm] |
|---|---|---|---|---|---|---|---|
| 2 | In-ACAC | 48 | 1.878 | 29 ± 1 | 115 ± 20 | 49 ± 3 | 100 ± 10 [4]<br>330 ± 20 |
| 4 | In-ACAC | ~ 1 | | 68 ± 3 | 280 ± 60 | | |
| 6 | In-CHO | 54 | 1.571 | 68 ± 3 | 165 ± 20 | 98 ± 6 | 280 ± 20 |
| 8 | In-CHO | ~ 51 | 1.522 | 69 ± 3 | 125 ± 20 | 91 ± 6 | 260 ± 20 |

**Table 1 Comparison of performances of differently formulated In-OLS measured in the quartz cell array.**

## 5 Conclusions and outlooks.

A novel original formulation to produce In-OLS based on In2MVA compound has been originally developed to produce In-OLS heavily indium doped (up to 80 g/l) having superior performances in term of light yield, light attenuation length, stability. These performances depends strongly on the formulation of the In2MVA compound and synthesis procedure, whose relevant parameters have been identified and kept under control. The developed In-OLS satisfy all the technical requirements to be used as target in a solar neutrino detector, whose feasibility depends on many other factors. The relevant outcome of this work is that the methodic developed to dope OLS with indium is easily extensible to dope OLS with other metals; as a matter of fact it has been adopted to produce Gd and Nd loaded OLS as well. Results will be presented in forthcoming papers.

## Acknowledgments

We are grateful to R. Raghavan and to C. Salvo for useful and stimulating discussion during all the work and to prof A. Bettini that stimulated the start up of the activity. A special thanks to all the LNGS chemistry service staff and in particular to G. Giusti, M. Balata and L. Ioannucci.
The work was supported by INFN – Commisione Scientifica Nazionale II and by Russian grants SS-5573.2006.2.

## Bibliography

1. R. Raghavan, Phys. Rev. Lett., **37** (1976) 259-262;
2. R. Raghavan, hep-ex/0106054;
3. C.Crieh (for LENS collaboration), Nucl. Phys. B (Proc. Suppl.) 168 (2007) 122-124
4. C. Buck at al, Journal of Radioanalytical and Nucl. Chem. 258 (2003) 255-263.
5. C. Buck at al, Journal of Luminescence 106 (2004) 57-67.
6. C. Buck at al, Nucl. Phys. B (Proc. Suppl.) 143 (2005) 487.
7. N. A. Danilov at al, RADIOCHEMISTRY, **45** (2003) 140-146.
8. N. A. Danilov at al, RADIOCHEMISTRY, **47** (2005) 487-494.
9. C. Cattadori at al, LNGS/EXP-06/04-September 2004
10. C. Cattadori at al, LNGS/EXP-07/04-September 2004
11. I. Barabanov at al, Nucl. Phys. B (Proc. Suppl.) **143** (2005) 559.

[2] 100 is the average for PXE cells
[3] Light attenuation length measured by spectrofotometry at 430 nm without addition of primary fluor.
[4] Light attenuation length measured by spectrofotometry at 430 nm with addition of primary fluor.